\documentclass{icrc2009}
\usepackage{amssymb}
\usepackage{graphicx}   % for including figures
\usepackage[caption=false]{caption}    % for captions
\usepackage[font=footnotesize]{subfig} % subfig.sty for a double column floating figure using two subfigures
\usepackage{fixltx2e}
\usepackage{url}

\newcommand{\shorttitle}[1]%
{\markboth{Proceedings of the 31\MakeLowercase{$^{st}$} ICRC, {\L}\'{o}d\'{z} 2009}{#1} }
\newcommand{\etal}{\MakeLowercase{\textit{et al.}}} % "et al."

\begin{document}
\title{Hadron-gamma discrimination from an orbital UHECR observatory}

\author{\IEEEauthorblockN{A.D. Supanitsky\IEEEauthorrefmark{1},
			  G. Medina-Tanco\IEEEauthorrefmark{1},
                          K. Asano\IEEEauthorrefmark{2},
                          D. Cline\IEEEauthorrefmark{4},
                          T. Ebisuzaki\IEEEauthorrefmark{5},\\
                          S. Inoue\IEEEauthorrefmark{6},
                          P. Lipari\IEEEauthorrefmark{7},
                          N. Sakaki\IEEEauthorrefmark{5},
                          A. Santangelo\IEEEauthorrefmark{8},
                          K. Shinozaki\IEEEauthorrefmark{5},
                          G. Sigl\IEEEauthorrefmark{9},
                          Y. Takahashi\IEEEauthorrefmark{10} \\ and 
                          M. Teshima\IEEEauthorrefmark{11} for the JEM-EUSO Collaboration
} \\

\IEEEauthorblockA{\IEEEauthorrefmark{1} Departamento de F\'isica de Altas Energ\'ias, Instituto de Ciencias 
Nucleares, Universidad Nacional Aut\'onoma \\ de M\'exico, A. P. 70-543, 04510, M\'exico, D. F., M\'exico.}
\IEEEauthorblockA{\IEEEauthorrefmark{2} Interactive Research Center of Science, Graduate School of Science, 
Tokyo Institute of Technology, \\ 2-12-1 Ookayama Meguro-ku Tokyo 152-8550, Japan.}
\IEEEauthorblockA{\IEEEauthorrefmark{4} Department of Physics and Astronomy, University of California, Los Angles, USA.}
\IEEEauthorblockA{\IEEEauthorrefmark{5} RIKEN Advanced Science Institute, Japan.}
\IEEEauthorblockA{\IEEEauthorrefmark{6} Dept. of Physics, Kyoto University, Kyoto 606-8502, Japan.}
\IEEEauthorblockA{\IEEEauthorrefmark{7} INFN-Roma La Sapienza, I-00185 Roma, Italy.}
\IEEEauthorblockA{\IEEEauthorrefmark{8} Institute fuer Astronomie und Astrophysik Kepler Center for Astro and 
Particle Physics\\ Eberhard Karls University Tuebingen Germany.}
\IEEEauthorblockA{\IEEEauthorrefmark{9} Institut theoretische Physik Universitaet Hamburg Luruper Chaussee 149 D-22761 Hamburg, Germany.}
\IEEEauthorblockA{\IEEEauthorrefmark{10} Dept. of Physics, The University of Alabama in Huntsville, Huntsville, AL35899, USA.}
\IEEEauthorblockA{\IEEEauthorrefmark{11} Max-Planck-Institut f\"ur Physik, F\"ohringer Ring 6, D-80805 M\"unchen, 
Germany.}

}

% please write the preseter's name and short title (3-4 words maximum)
%    which will appear at the header of the even pages.
\shorttitle{A.~D.~Supanitsky \etal Hadron-gamma discrimination}
\maketitle

\begin{abstract}

The identification of very high energy photons is of great importance for the understanding of the origin of 
extreme energy cosmic rays (EECR). Several can be the sources of high energy photons at Earth. A guaranteed 
component is the flux of high energy photons expected as a consequence of the interaction of cosmic rays with 
the cosmic photon background. Another contribution may be expected as by-product at the acceleration sites of 
protons and nuclei, although such flux should be strongly suppressed for distant sources. On the other hand, 
top-down scenarios involving the decay of super heavy relic particles or topological defects, even if not 
currently favored, have as a characteristic signature an increasingly dominant flux of photons at the highest 
energies. In this work we study the statistical separation between hadron and photon showers at energies where 
both, LPM effect and magnetospheric interactions are important for the development of the cascades. We consider 
a detector with the same orbital characteristics as JEM-EUSO, but disregard trigger and reconstruction efficiencies, 
in order to define the maximum ideal discrimination power attainable. 

\end{abstract}


\section{Introduction}

The cosmic ray energy spectrum must have at least a minor component of ultra high energy photons. 
This component may receive contributions from different sources. Besides the expected flux generated 
by the propagation of the EECR in the intergalactic medium \cite{Gelmini:07}, photons may also be 
originated in different astrophysical environments as by-products of particle acceleration in 
nearby cosmic ray sources (e.g. \cite{CentaurusA:08}) and, fundamentally, in top-down scenarios 
involving the decay of super heavy relic particles or topological defects \cite{Aloisio:04}. Extreme 
energy photons have not been unambiguously observed yet. However, it is expected that JEM-EUSO 
\cite{Ebisuzaki:09,Gustavo:09}, with its unprecedented exposure will change this situation in the 
next few years.

In the present work we estimate the photon-proton discrimination power of JEM-EUSO. 
In particular we develop two complementary techniques to evaluate an upper limit on the fraction of 
photons relative to proton primaries in the integral cosmic rays flux by using the atmospheric depth 
of maximum development, $X_{max}$, of the corresponding atmospheric showers. The longitudinal 
evolution of high energy photon showers is dominated by the interplay between magnetospheric 
photon splitting and the LPM effect. The first process is highly dependent on the incoming direction 
of the photon with respect to the geomagnetic field and its intensity. Therefore, the translation of 
JEM-EUSO along its orbit is a distinctive parameter which adds richness and complexity to the 
analysis with respect to a traditional Earth bound observatory and must be taken into account.

\section{Numerical Approach}

Given a sample of $N$ events, an ideal upper limit to the photon fraction may be calculated under 
the a priori assumption that actually no photon exists in the sample:
\begin{equation}
\mathcal{F}_{\gamma}^{min} = 1-(1-\alpha)^{1/N}
\label{Fuplideal}
\end{equation} 
where $\alpha$ is the confidence level of rejection. However, in practice, the probability of the 
existence of photons must be realistically assessed through some observational technique which 
involves the determination of experimental parameters, which leads unavoidably to less restrictive 
upper limits than the previous one.

In this work the $X_{max}$ parameter is considered for the discrimination of protons and photons showers. 
A shower library was generated by using the program CONEX \cite{conex} which consist of $3 \times 10^{5}$ 
proton showers following a power law energy spectrum of spectral index $\gamma = -2.7$ in the interval 
[$10^{19.8}, 10^{21}$] eV and with uniformly distributed arrival directions. Also $> 5.1 \times 10^{5}$ 
photon showers were generated under the same conditions but in this case cores were also uniformly distributed 
on the surface of the Earth in order to properly take into account pre-showering (i.e., photon splitting) in 
the geomagnetic field.    

A measure of the discrimination power of the $X_{max}$ parameter is given by the merit factor,
\begin{equation}
\eta = \frac{\textrm{med}[X_{max}^\gamma]-\textrm{med}[X_{max}^{pr}]}{\sqrt{(\Delta X_{max}^\gamma)^2+(\Delta X_{max}^{pr})^2}}
\label{eatdef}
\end{equation}
where $\textrm{med}[X_{max}^A]$ is the median of $X_{max}^A$ ($A=\gamma, pr$) distribution and $\Delta X_{max}^A$ is one 
half of the length of the region of 68\% of probability of $X_{max}^A$ distribution. 

\begin{figure}[!t]
\centering
\includegraphics[width=3.2in]{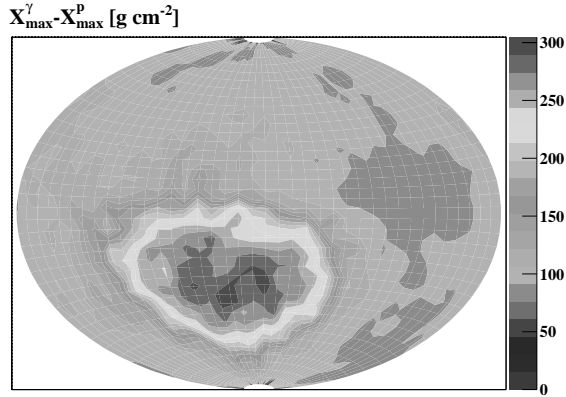}
\caption{Contour plot of $X_{max}^\gamma-X_{max}^{pr}$ as a function of latitude and longitude in the Earth. The 
showers considered are such that $\theta \in [30^\circ, 60^\circ]$ and $E \in [10^{19.8}, 10^{20}]$ eV.}
\label{Map}
\end{figure}

Fig. \ref{Map} shows a contour plot of $X_{max}^\gamma-X_{max}^{pr}$ as a function of latitude and longitude
of the core on the Earth, for $\theta \in [30^\circ, 60^\circ]$ and $E \in [10^{19.8}, 10^{20}]$ eV. It can be seen that 
there are regions over the Earth surface where this difference is larger and in which the discrimination between
protons and photons is more efficient.

\begin{figure}[!t]
\centering
\includegraphics[width=3.3in]{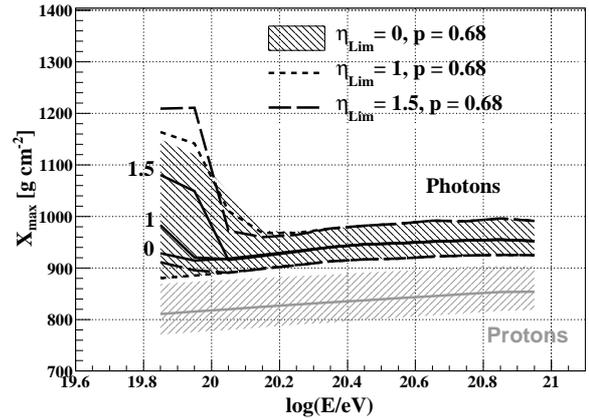}
\caption{Median and region of 68\% of probability of $X_{max}$ as a function of primary energy for $\theta \in [30^\circ, 60^\circ]$.
In the case of photons different masks are considered, $\eta_{Lim}=0$ (all the events), $\eta_{Lim}=1$ and $\eta_{Lim}=1.5$.}
\label{XmaxCL}
\end{figure}

Motivated by this result the concept of mask, $\Omega(\eta_{Lim})$, is introduced as those regions over the Earth surface where 
$\eta$ is larger than a given value $\eta_{Lim}$. Fig. \ref{XmaxCL} shows the median of $X_{max}$ 
and the region of 68\% of probability as a function of primary energy for protons and photons with
$\theta \in [30^\circ, 60^\circ]$. For the case of photons different masks are considered, $\Omega(\eta_{Lim}=0)$ (all events), 
$\Omega(\eta_{Lim}=1)$ and $\Omega(\eta_{Lim}=1.5)$. Note that the masks are functions of primary energy. From the figure it can be 
seen that for the photons there are two well defined regions, the first corresponds to primary energies smaller than $10^{20.1}$ eV 
in which the $X_{max}^\gamma$ distribution is composed by LPM dominated showers that both, suffer and do not suffer photon splitting 
in the Earth magnetic field. In the second region all showers undergo photon splitting which generates, on average, smaller values of 
$X_{max}^\gamma$ and smaller fluctuations \cite{Vankov:03}.      

\begin{figure}[!t]
\centering
\includegraphics[width=3.2in]{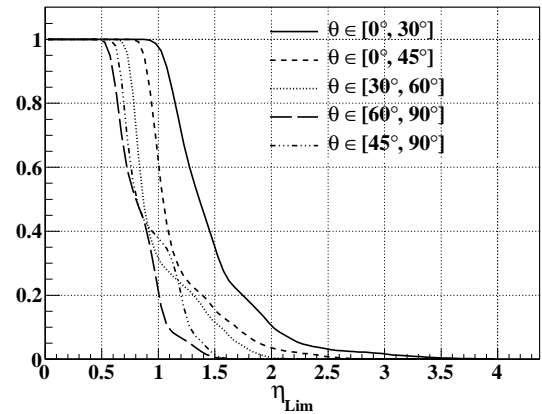}
\caption{Fraction of events as a function of $\eta_{Lim}$ for different cuts in zenith angle and for 
$E\in[10^{19.8},10^{20}]$ eV.}
\label{EtaLim}
\end{figure}

The $X_{max}^\gamma$ distributions obtained for masks with larger $\eta_{Lim}$ allow a better separation between protons and photons.
However, the total number of events also depends on the assumed mask and, in particular, decreases with $\eta_{Lim}$. Fig. \ref{EtaLim}
shows the fraction of events as a function of $\eta_{lim}$ for different cuts in zenith angle and for $E\in[10^{19.8},10^{20}]$ eV. 
It can be seen that for vertical showers larger values of $\eta$ are obtained which are also distributed over a wider range.     

Two methods were developed in order to calculate an upper limit for the photon fraction. The first one is based on the abundance 
estimator first introduced in \cite{Supanitsky:08},
\begin{equation}
\xi_{X_{max}} = \frac{1}{N} \sum_{i=0}^{N} \frac{f_{\gamma}(X_{max}^i)}{f_{\gamma}(X_{max}^i)+f_{pr}(X_{max}^i)}
\label{eatdef}
\end{equation}
where $f_\gamma(X_{max})$ and $f_{pr}(X_{max})$ are the photon and proton distribution functions, $X_{max}^{i}$ are experimental values 
of $X_{max}$ and $N$ is the sample size. $\xi_{X_{max}}$ is an estimator of the photon abundance, $c_{\gamma}=N_\gamma/N$ where $N_\gamma$ 
is the number of photons in the sample. The mean value of $\xi_{X_{max}}$ is a linear function of $c_{\gamma}$, its standard deviation is 
proportional to $1/\sqrt{N}$ and, for large values of $N$, it follows a Gaussian distribution.

For the case in which $\xi_{X_{max}}$ is compatible with a pure proton sample, an upper 
limit to the photon fraction, $c_\gamma^{min}$, can be obtained from,
\begin{equation}
c_\gamma^{min}=\frac{4}{u_1^2} \left( \frac{u_1 \sqrt{v_2}}{\sqrt{N}} + \frac{v_1}{N} \right)
\label{cgmin}
\end{equation}
where $u_1=\alpha_1-\alpha_2$, $u_2=\alpha_2$, $v_1=\alpha_3-\alpha_4+\alpha_2^2-\alpha_1^2$ and $v_2=\alpha_4-\alpha_2^2$. Here 
\begin{eqnarray}
\alpha_1 &=& \int dX_{max} \frac{f_{\gamma}(X_{max})^2}{f_{\gamma}(X_{max})+f_{pr}(X_{max})}, \\
\alpha_2 &=& \int dX_{max} \frac{f_{\gamma}(X_{max}) f_{pr}(X_{max})}{f_{\gamma}(X_{max})+f_{pr}(X_{max})}, \\
\alpha_3 &=& \int dX_{max} \frac{f_{\gamma}(X_{max})^3}{[f_{\gamma}(X_{max})+f_{pr}(X_{max})]^2}, \\
\alpha_3 &=& \int dX_{max} \frac{f_{\gamma}(X_{max})^2 f_{pr}(X_{max})}{[f_{\gamma}(X_{max})+f_{pr}(X_{max})]^2}. 
\end{eqnarray}

The distribution functions needed to calculate $c_\gamma^{min}$ are obtained from the simulated data by using 
the non-parametric method of kernel superposition with adaptive bandwidth \cite{Silvermann:86}.

\begin{figure}[!t]
\centering
\includegraphics[width=3.2in]{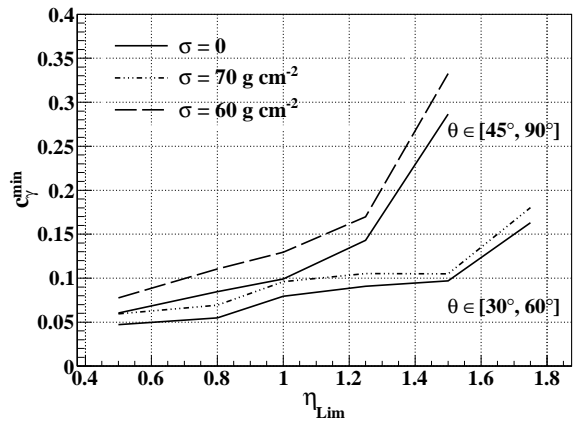}
\caption{$c_\gamma^{min}$ as a function of $\eta_{Lim}$ for $E\in[10^{19.8},10^{20}]$ eV, $\theta \in [30^\circ, 60^\circ]$ 
and $\theta \in [45^\circ, 90^\circ]$ and with and without assuming a Gaussian uncertainty of $70$ g cm$^{-2}$
for $\theta \in [30^\circ, 60^\circ]$ and $60$ g cm$^{-2}$ for $\theta \in [45^\circ, 90^\circ]$.}
\label{Cgmin}
\end{figure}

It can be seen from Eq. (\ref{cgmin}) that, the larger the sample size, the smaller $c_\gamma^{min}$. Although, it
is not obvious from this expression, it is possible to show that for larger values of $\eta$ also smaller 
values of $c_\gamma^{min}$ are obtained. Fig. \ref{Cgmin} shows $c_\gamma^{min}$ as a function of $\eta_{Lim}$,
i.e. for different masks, for $E\in[10^{19.8},10^{20}]$ eV, $\theta \in [30^\circ, 60^\circ]$ and 
$\theta \in [45^\circ, 90^\circ]$ and with and without assuming a Gaussian uncertainty of $70$ g cm$^{-2}$
for $\theta \in [30^\circ, 60^\circ]$ and $60$ g cm$^{-2}$ for $\theta \in [45^\circ, 90^\circ]$. It can
be seen that $c_\gamma^{min}$ increases with $\eta_{Lim}$, which means that although the discriminator
power of $X_{max}$ increases the number of events decreases so rapidly producing larger values of 
$c_\gamma^{min}$, i.e. in this case the number of events is more important than the discrimination
power of $X_{max}$ for a given mask. 

The second method developed here consists in finding a cut on $X_{max}$ which minimizes the expression of the 
upper limit obtained assuming a pure proton composition, 
\begin{equation}
\mathcal{F}_{UL}(X_{max}^c) = \frac{N_\alpha(N F_{pr}(X_{max}^c))}{N F_{\gamma}(X_{max}^c)},
\label{Fuplimcut}
\end{equation}
where $N_\alpha(n)$ is the upper limit on the number of photons $n$ at a confidence level $\alpha$ obtained assuming a Poisson 
distribution and  
\begin{equation}
F_A(X_{max}^c)=\int_{X_{max}^c}^\infty d X_{max}\ f_A(X_{max}).
\end{equation}

\begin{figure}[!t]
\centering
\includegraphics[width=3.2in]{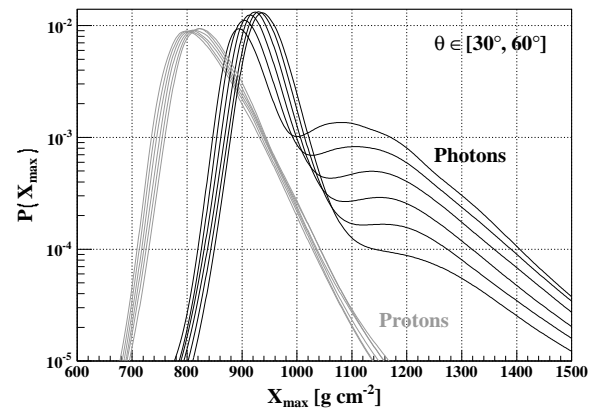}
\caption{Estimates of the protons and photons distribution functions of $X_{max}$ for threshold energies from 
$10^{19.8}$ eV to $10^{20.3}$ eV in steps of $\log(E/\textrm{eV})=0.1$ and for $\theta \in [30^\circ, 60^\circ]$.
A power law energy spectrum of spectral index $\gamma = -4.2$ is assumed. The mean value of the distributions 
increases with primary energy.}
\label{PDFEth}
\end{figure}

In order to study the upper limit for a given threshold energy the distribution functions of $X_{max}$ for protons 
and photons are obtained from MC data, by using the non-parametric method of kernel superposition mentioned
above. Fig. \ref{PDFEth} shows the estimates of the proton and photon distribution functions for 
$\theta \in [30^\circ, 60^\circ]$ and for threshold energies from $10^{19.8}$ eV to $10^{20.3}$ eV in steps of 
$\log(E/\textrm{eV})=0.1$. A power law energy spectrum of spectral index $\gamma = -4.2$ following the shape of the 
Auger spectrum \cite{AugerSpec:08} is assumed. The figure shows, as expected, that as the threshold energy increases 
the bump due to the photons that do not suffer photon splitting in the geomagnetic field becomes progressively less 
important.   

\begin{figure}[!t]
\centering
\includegraphics[width=3.2in]{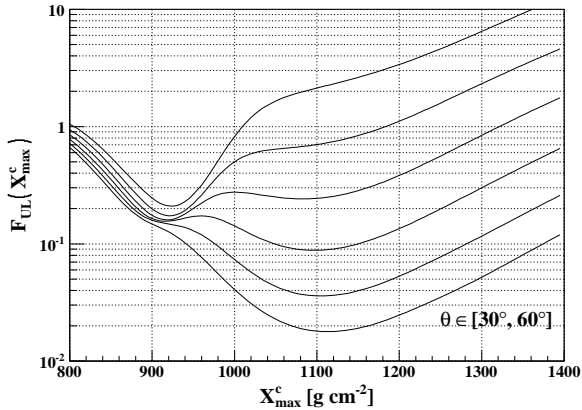}
\caption{Upper limit as a function of $X_{max}^c$ for threshold energies from $10^{19.8}$ eV to $10^{20.3}$ eV 
in steps of $\log(E/\textrm{eV})=0.1$ and for $\theta \in [30^\circ, 60^\circ]$.}
\label{ULcut}
\end{figure}

Fig. \ref{ULcut} shows $\mathcal{F}_{UL}(X_{max}^c)$ as a function of $X_{max}^c$ obtained by using the distribution 
functions of Fig. \ref{PDFEth}, for $\alpha=0.95$ and for a total number of events above $10^{19.8}$ eV $N=2250$ (4500 
events in total but half of them have $\theta \in [30^\circ, 60^\circ]$). It can be seen that $\mathcal{F}_{UL}(X_{max}^c)$ 
reaches a minimum which depends on the threshold energy. Note that there is a transition at $E_{th}=10^{20}$ eV from which 
the minimum is reached at larger values of $X_{max}^c$. This is due to the change in the shape of the photon distribution 
function when the threshold energy increases (see Fig. \ref{PDFEth}). Finally, the upper limit is obtained evaluating 
$\mathcal{F}_{UL}$ in $X_{max}^c$ of the minimum.    

\begin{figure}[!t]
\centering
\includegraphics[width=3.3in]{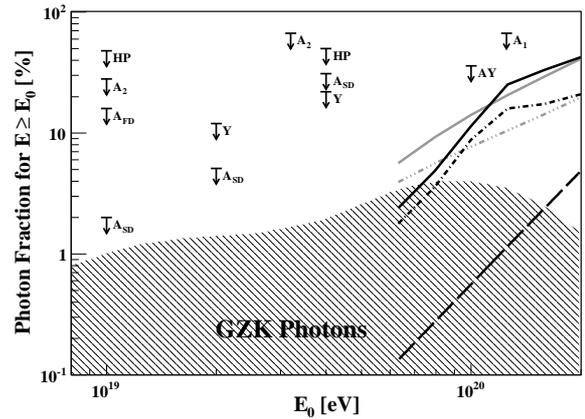}
\caption{The upper limits on the fraction of photons in the integral cosmic ray flux at 95\% of confidence level. 
Dashed line corresponds to the ideal case in which it is known that there is no photons in the data. Solid and 
dash-three dots-dash gray lines are the upper limits obtained by using $\xi_{X_{max}}$ method assuming a $70$ 
g cm$^{-2}$ of Gaussian uncertainty and no uncertainty in the determination of $X_{max}$, respectively. Solid and 
dash-dot-dash black lines are the upper limits obtained by using the optimized cut method assuming a $70$ 
g cm$^{-2}$ of Gaussian uncertainty and no uncertainty in the determination of $X_{max}$, respectively. Shadow 
region is the prediction for the GZK photons \cite{Gelmini:07}. Black arrows are experimental limits, 
HP: Haverah Park \cite{Ave}; A$_1$, A$_2$: AGASA \cite{Risse:05,Shinozaki:02}; A$_{\textrm{FD}}$, 
A$_{\textrm{SD}}$: Auger \cite{augerFD,augerSD}; AY: AGASA-Yakutsk \cite{Rubtsov:06}; 
Y: Yakutsk \cite{Glushkov:07}.}
\label{ULFinal}
\end{figure}

Fig. \ref{ULFinal} shows the upper limits on the fraction of photons in the integral cosmic ray flux, at 95\%
of confidence level, obtained in the ideal case $\mathcal{F}_{\gamma}^{min}$ (dashed line), by using the 
$\xi_{X_{max}}$ method for $\theta \in [30^\circ, 60^\circ]$ and assuming a $70$ g cm$^{-2}$ of Gaussian 
uncertainty and no uncertainty in the determination of $X_{max}$ (solid and dash-three dots-dash gray lines, 
respectively), by using the optimized cut method for $\theta \in [30^\circ, 60^\circ]$ and assuming a 
$0$ and $70$ g cm$^{-2}$ of Gaussian uncertainty in the determination of $X_{max}$ (dash-dot-dash and solid 
black lines, respectively) and the upper limits obtained by different experiments. It can 
be seen that the upper limits obtained by using both methods introduced here are about one order of magnitude 
larger than the ideal case. This is due to the limitation imposed by the $X_{max}$ parameter to discriminate 
between photons and protons. Moreover, for the energies $>10^{20}$ eV, the $X_{max}$ distribution of 
photon showers is dominated by photon splitting decreasing the discrimination power 
of this parameter.

Fig. \ref{ULFinal} also shows that for energies bellow $10^{20.1}$ eV the optimized cut method results
better than $\xi_{X_{max}}$ method. This happens because the former takes advantage of the part of the 
distribution function originated by photons that do not suffer photon splitting. For larger energies the 
$\xi_{X_{max}}$ method results better because the part of the $X_{max}$ distribution of unconverted photons 
is less important and this methods takes into account the whole shape of proton and photon distribution
functions.  

\section{Conclusions}

In the present work we demonstrate that, based on pure statistics, JEM-EUSO will be able to set an upper limit 
to the photon fraction at the highest energies of $\sim 10^{-1}$ \%, well inside the GZK-photon flux expectation. 
A more realistic estimate, taking into account a conservative discrimination parameter and its experimental 
uncertainty the upper limit is larger but still considerably smaller than the corresponding values set so far by
existing experiments. Therefore, given its large exposure, JEM-EUSO should be the first cosmic ray experiment 
to unambiguously detect cosmogenic photons.

\end{document}